# Why Decussate? Topological Constraints on 3D Wiring

**TROY SHINBROT**[1]* **AND WISE YOUNG**[2]

[1]Department of Biomedical Engineering, Rutgers University, Piscataway, New Jersey
[2]Department of Cell Biology and Neuroscience, Rutgers University, Piscataway, New Jersey

## ABSTRACT

Many vertebrate motor and sensory systems "decussate" or cross the midline to the opposite side of the body. The successful crossing of millions of axons during development requires a complex of tightly controlled regulatory processes. Because these processes have evolved in many distinct systems and organisms, it seems reasonable to presume that decussation confers a significant functional advantage—yet if this is so, the nature of this advantage is not understood. In this article, we examine constraints imposed by topology on the ways that a three-dimensional processor and environment can be wired together in a continuous, somatotopic, way. We show that as the number of wiring connections grows, decussated arrangements become overwhelmingly more robust against wiring errors than seemingly simpler same-sided wiring schemes. These results provide a predictive approach for understanding how 3D networks must be wired if they are to be robust, and therefore have implications both for future large-scale computational networks and for complex biomedical devices.   Anat Rec, 291:1278–1292, 2008.   © 2008 Wiley-Liss, Inc.

Key words: decussation; neuronal wiring; topology; dimensionality

## INTRODUCTION

In the human, over half of motor and sensory tracts decussate: for example, the somatosensory, corticospinal, and corticorubral tracts all cross over to the opposite side of the body. Decussation predominantly occurs in tracts that mediate spatially organized data, whereas nonsomatotopic and phyllogenetically older tracts do not decussate: examples include the olfactory, vestibulospinal, and reticulospinal systems (Vulliemoz et al., 2005).

A large body of research into the precise anatomy of decussating neuronal pathways has been established, beginning with Hall in 1884 (Wollaston, 1824; Hall and Hartwell, 1884; Marshall, 1936), and noted as early as 1709 by Mistichelli (Mistichelli, 1709). More recently, developments in genomic and molecular techniques have allowed researchers to investigate detailed mechanisms by which these decussating pathways develop (Jouet et al., 1994; Letinic and Rakic, 2001). Some of the molecules responsible for guiding the decussation of neural pathways are evolutionarily quite primitive, for example, netrin and its receptors UNC-5 and UNC-40 are present in *C. elegans* (Hedgecock et al., 1990).

From this body of developing knowledge, it has become clear that the mechanisms involved in successfully navigating the crossing of tracts that contain many millions of axons are extremely intricate and precise. Yet it remains uncertain what practical or evolutionary function these difficult maneuvers serve. Classical explanations suggest that crossings originated in the optical chiasm because of a need for organisms to reconcile both left and right hemispheres of view. According to this reasoning, the crossing of visual pathways necessitates a crossing of motor pathways, especially during primitive escape reflexes of limbed animals that must

Grant sponsor: New Jersey Commission on Higher Education.

*Correspondence to: Troy Shinbrot, Department of Biomedical Engineering, Rutgers University, Piscataway, NJ 08854.
E-mail: shinbrot@rutgers.edu

Received 12 June 2007; Accepted 19 February 2008

DOI 10.1002/ar.20731

Published online in Wiley InterScience (www.interscience.wiley.com).

push away from a dangerous stimulus [see (Vulliemoz et al., 2005) for a historical review].

Notwithstanding this explanation, the functional consequences of failure to decussate in various developmental disorders are subtle and incompletely understood (Ferland et al., 2004; Jouet et al., 1994; Vits et al., 1994; Vulliemoz et al., 2005), and numerous basic scientific questions remain unanswered. For example, older pathways that are involved in escape reflexes and that have been present throughout the evolution of vision (e.g., the lateral vestibulospinal tract) remain largely undecussated both in limbed and nonlimbed animals. On the other hand, *C. elegans*, with no limbs and only the most elementary of visual apparatus, has evolved a decussated ventral tract, and blind moles have retained decussated motor as well as sensory pathways (Catania and Kaas, 1997). Moreover, pathways that have little to do with escape, especially those involving fine motor movement, are largely decussated (e.g., the human lateral corticospinal tract). Thus, much remains to be understood about the origins of decussation.

In this article, we address the question of why the intricate developmental process of decussation has evolved in so many different systems and organisms from a topological perspective that have not, to our knowledge, been considered before. This topological approach, seemingly of mere academic interest, turns out to have potential for significant practical consequences that we briefly outline.

In the Effect of Three-Dimensionality section, we describe a topological cause of crossings in tracts involving the mapping of 3D information. We emphasize that this cause persists when one needs to map a 2D surface (e.g., the skin) embedded in 3D. We show using explicit examples that wiring schemes that lack crossings in 3D systems necessarily involve spatial processing deficits. Next, in the Alternative Crossing Strategies section, we examine alternative wiring schemes that may reduce the total number of crossings, and then in the Analysis section, we compare the numbers of crossings present in these alternative schemes. We calculate how the numbers of crossings change as the system size grows, and we provide an evolutionary simulation that attempts to minimize crossing numbers through random mutations. Next, in the Discussion section, we describe both the mathematical findings and their possible physiological interpretations, and we discuss some comparative examples in different mammals. Finally, in the Conclusions section, we summarize possible implications for developmental and regenerative therapies.

## EFFECTS OF THREE-DIMENSIONALITY

We begin with the self-evident observation that crossed connections seem to be unnecessary. As depicted in cartoon form in Fig. 1a, an idealized 2D "brain" can be connected point-to-point with its counterpart "body" motor or sensory system without any need to cross connections. On the other hand, brain and body structures reside in three-dimensions (3D) and the situation changes markedly for 3D networks (Changeux, 1985; Snyder et al., 1998; Graziano et al., 2002). To see this, consider the simplest excursion into three-dimensionality: that which results when flat sheet is folded out of the plane. This is indicated by the arrows and the resulting caricature shown in Fig. 1b. We first examine the case shown, in which both brain and body are folded in the same orientation; we consider another folding orientation shortly.

Folds in the nervous system are a topic unto themselves: folds are ubiquitous in the mammalian cerebellum and in the primate cerebral cortex, and folding abnormalities are implicated in developmental disorders including autism (Hardan et al., 2004) and polymicrogyria (Piven et al., 1990; Piao et al., 2004). The effects of folds *per se* are beyond the scope of this article; for our purposes, we focus on the observation that as Fig. 1b shows, when the connections travel through a central column, crossings can appear from no cause beyond the act of folding of a sheet into 3D.

We emphasize that the act of folding, and the consequent crossings of connections, shown in Fig. 1 has no bearing on the ability of a 2D brain and body to communicate. Two sheets can be wired point-to-point and then folded to any degree of complexity without any loss of information. However, the central point that we will demonstrate is that these folds and crossings are crucial to the ability to convey 3D information.

The distinction between the 2D sheets of the cartoon brain and body and the 3D environment in which the body resides (and which the brain reconstructs as a stereognostic model) is not obvious, and a goal of this article is to analyze this distinction in some detail. We begin by analyzing the relevance of the crossings that result from the fold shown in Fig. 1b.

## Crossings

The cause of the crossing shown in Fig. 1b is not mysterious: it is merely the fact that once point A in the figure has been connected to its counterpart, A', point B beneath cannot be connected to B' in the symmetric way shown on the right of the figure without crossing the A-A' connection.

Crossings are important to neuronal function for three reasons. One, the number of crossings ought to be minimized because each crossing takes up space and adds lengths to connections (Koulakov and Chklovskii, 2001; Chklovskii et al., 2002; Chklovskii and Koulakov, 2004), both of which violate Cajal's laws of neuronal optimization (Cajal, 1995; Llinas, 2003). We recall for the reader that these laws require that neuronal arrangements (1) minimize space, (2) maintain a necessary size, and (3) minimize conduction time. Two, in order for axons to target accurately during development or following regenerative therapy, crossings must be (and are) (Yoshikawa et al., 2003) tightly regulated. Put in simplistic terms, if a motor axon destined for the foot changes places, i.e., crosses inappropriately, with an axon destined for the hand, either motor function will be compromised or the crossing will need to be untangled in the brain before constructing the primary motor homunculus. And three, information is required to prescribe how each crossing must occur in order that crossings be minimized and accurately regulated. As we will show, some crossing strategies require more bits of information than there are active genes in the human genome, whereas others (especially decussation) require dramatically less information to achieve accurately.

We digress to clarify a mathematical point: the "crossing" shown in Fig. 1b obviously depends on the projection shown in the cartoon; however, crossings can be unambiguously defined. One way to do so is to choose one fixed projection and mirror-reflect connections across a plane parallel to the projection plane, as shown in Fig. 1c. In this way, the original connections combined with their reflections define an inseparable "link" if and only if the connections cross in that projection (Bauer et al., 1980). Also at this point, we remark that crossings can appear if only the body is extended into 3D, an illustration of such a situation is shown in Fig. 1d. Thus, in the discussions that follow, one can consider the cartoon "brain" to literally be a portion of the 2D cortex (e.g., the primary somatosensory cortex), or one can view the cartoon brain as being the stereognostic representation of the body's environment (Kandel et al., 1991). Rather than draw both, in our discussions we sketch only the latter **representation**, because in this case both the body and the brain can be depicted as 3D objects, which makes analysis and discussion more straightforward than in the former, mixed 2D/3D representation. The reader should, however, remain aware that except where otherwise stated, the 3D "brain" that we sketch represents the stereognostic reconstruction of the body's environment, and to arrive at this reconstruction the brain uses both 2D processing systems (e.g., the somatosensory cortex) and 3D ones [e.g., the VPLN of the thalamus (Kandel et al., 1991)]. We clarify this distinction in concrete examples shortly.

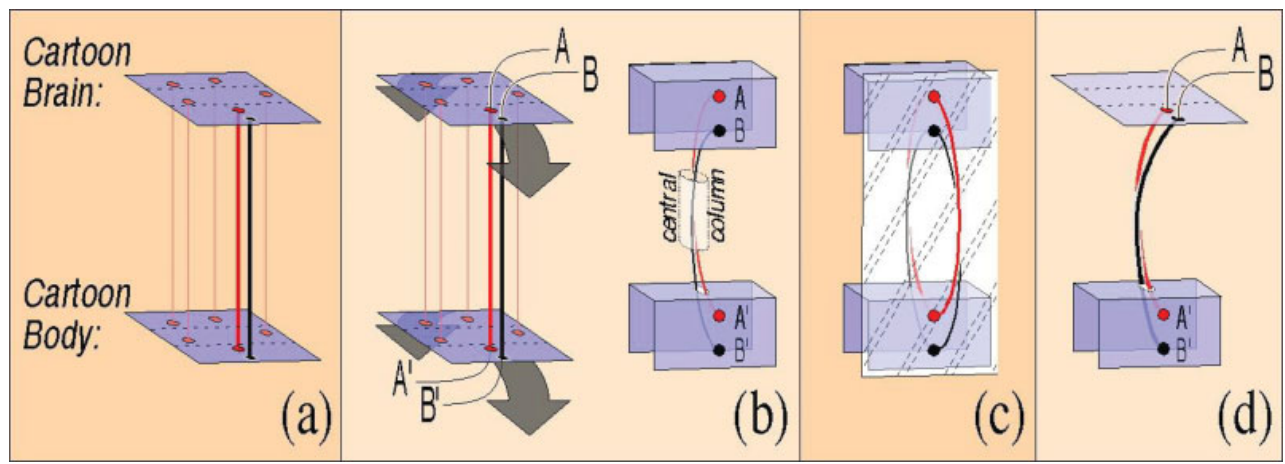

Fig. 1. Cartoon depictions of connections between a "brain" and a "body" in (a) 2D and (b) when folded into 3D. (a) For 2D brain and body, elements can be connected point-to-point without crossings. (b) Once brain or body extend into 3D in the orientation shown, connections that travel through a central column must cross. (c) Crossings of connections are formally defined to occur if a mirror reflection about a symmetry plane (hatched) generates an inseparable link as described in text. (d) Crossings can appear if only the body is extended into 3D as well.

## Why Do Crossings Occur at All?

In light of the observations that crossings ought to be minimized, both in principle and to reduce genetic information, we raise a very basic question: why does the central nervous system (CNS) employ crossings at all? For example, the configuration shown in Fig. 2 eliminates crossings altogether. In this configuration, the body and brain cartoons are folded in opposite directions, and consequently the positions of points A′ and B′ in Fig. 2b are reflected top-to-bottom with respect to their somatotopic images, A and B. On the other hand, images of points that are separated horizontally are not reflected. This implies that there must be a singularity in 3D representations of space along the bold green lines shown in Fig. 2b. A singularity is mathematically defined to be a point at which a measurable property diverges or becomes ill-defined. On the green lines of Fig. 2b, we say that the somatotopic mappings are singular in that on these lines, the 3D representations do not have a single defined value. On these singular lines, both an upright somatotopic image (in the horizontal plane) and a reversed one (in the vertical plane) coexist. This means that in order to track an ant crawling along the skin, the brain would need to reverse its somatotopic map as the ant crawled up one's chest (vertical and therefore reflected) and then over one's shoulder (horizontal and therefore upright). This presents an absurd state of affairs to any control or sensation network.

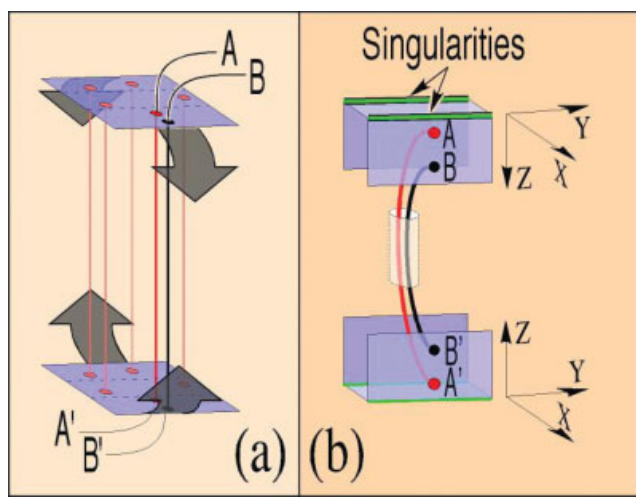

Fig. 2. (a) Cartoon of alternative construction in which brain and body surfaces are folded in opposite directions. (b) Crossings are now eliminated, but the vertical coordinate of the body and its percept are reversed, whereas the horizontal coordinates remain upright: note that the upright coordinates of the "body" are shown as (X, Y, Z), whereas those of its representation in the "brain" are (X, Y, −Z). Crucially, this produces geometric "singularities" (green) in somatotopic mapping, where the upright horizontal plane meets the reversed vertical plane, as described in the text.

This is a peculiar assertion that may be difficult to visualize. To demonstrate this effect, we therefore consider two examples in the configuration shown in Fig. 2. First, we examine how 3D somatotopic sensation would appear in such a configuration, and second, we describe how 3D motor function would change without crossings.

***Example 1: 3D sensory somatotopy without crossings is intractably complicated.*** As a first example in Fig. 3a, we sketch a finger wired without crossings beneath a caricature of the brain. Light touch receptors on a finger lie on a single dermatome on a 2D sheet; likewise the location of the finger in the primary somatosensory homunculus lies on a 2D sheet, so that this is as exact as one can make such a representation. In keeping with our earlier remarks, for simplicity of discussion, we sketch the brain as a 3D stereognostic reconstruction of the finger's environment and not as the intermediate 2D somatosensory cortex, but the reader can confirm that in this case, the results that we describe would apply equally if the sketches illustrated the "brain" as a single flat sheet.

If the system shown is connected without crossings, a point $(X, Y, Z)_{body}$ in the finger's environment must become mapped to a point $(X, Y, -Z)_{brain}$ in the stereognostic percept in the brain. This dictates that horizontal positions $(X, Y)_{body}$ are mapped upright to $(X, Y)_{brain}$, but vertical positions $(X, Z)_{body}$ are inverted to $(X, -Z)_{brain}$. The negative sign in the vertical mapping (and its lack in the horizontal one) causes the finger's percept to be intractably complicated.

To make the problem concrete, in Fig. 3, we sketch a finger and its percept as they would appear if the finger were wired without crossings, as in Fig. 2b. In Fig. 3a, we indicate the finger facing forward and then turned to face downward; in both cases, we depict the finger's unchanging 3D environment as a cube. What we will show is that if the finger is wired without crossings, then its percept becomes unexpectedly difficult to analyze if it is subjected to even such a simple operation as the rotation indicated by Fig. 3a. The finger is shown schematically along with its percept in Fig. 3b,c for the forward and downward facing cases respectively. In both cases, the environment is unchanged, which we indicate by sketching identical lettering: "top," "front," and "bottom" on the sides of a 3D cube surrounding the finger. We caution that in this case, the upper cube represents the brain's percept, the finger represents a part of the body, and the lower cube now depicts the finger's environment, and not a body part.

If we first consider the percept of the finger's environment in the orientation of Fig. 3b, we see that the lettering "front" appears to be flipped top-to-bottom, whereas the lettering "top" and "bottom" are correctly oriented. The red and black spots shown in Fig. 3 correspond identically to those of Fig. 2b, and the flipping of the front surface is a necessary result of reversing the red and black spots, as was shown in Fig. 2b to occur when crossings are removed. If we now consider Fig. 3c, we find that in its percept, the lettering "bottom," is flipped, but the lettering "front" is now upright. In a finger wired without crossings as shown, the environment facing the finger pad must be perceived fundamentally differently from the environment facing the finger sides, and this implies that in order to interpret what the finger is touching, the direction that it is facing in must be known. Consequently as the finger is rotated, the percept of an unchanging environment must undergo a continuously changing inversion. Tasks such as reading Braille with the pad or side of the finger, identifying the direction on the body an insect might crawl along, and producing consistent percepts of multiple differently facing body parts would be prohibitively complicated in any but the simplest organism wired in this way. We reiterate that this problem does not arise if crossings of connections are permitted, for example, as in Fig. 1.

***Example 2: 3D motor function without crossings is slow and uncoordinated.*** The significance of the singularities shown in Fig. 2b between vertical and horizontal representations in a system wired without crossings can be illustrated more tangibly in a second example. Consider the control problem shown in Fig. 4, possibly familiar to the reader as a child's game. Here, we depict moving our left index finger in a clockwise direction (as viewed from the right). It is manifestly trivial to simultaneously move the right index finger in the same direction; however, it is markedly more difficult to move the right index finger in the opposite direction (counterclockwise viewed from the right). This problem is intrinsically associated with the brain's 3D representation of space: the difficulty is not present if the fingers are pointed forward: one can sketch forward-facing circles in either direction and with any speed with either hand (Bogaerts and Swinnen, 2001). This is so because there are singularities in the motion shown in Fig. 4 that vanish when both fingers point forward.

To identify these singularities, we dissect the motion of Fig. 4. With practice, one can perform the counterrotating motion shown by using the strategy described to the right of the figure, in which each hand circumscribes a square in space. However, to employ this, or any other effective strategy, one must at some singular point (the corners of the squares here) simultaneously move the arms in-phase in one direction (front and back in this example) and anti-phase in a different direction (up and down here) (Salter et al., 2004; Swinnen and Wenderoth, 2004). This singularity appears because the fingers are pointed in opposite directions: if one fingertip (pointing in the $+X$-direction) is represented as $(X, Y, Z)$, the other fingertip (pointing in the $-X$-direction) would be represented as $(-X, Y, Z)$. When both fingers face forward, left and right hands share a single coordinate system, and so the singularities, and the associated motor control problems, vanish.

This example provides us with an additional result as well, namely that without crossings, there appears to be a loss of speed and control flexibility. This can be made evident by practicing the procedure described in Fig. 4 and then repeating the same procedure, but where the fingers cross at top and bottom of the same circumscribed squares (instead of front and back as previously). Achieving the first motion is of little help in learning the second, and similarly for a third motion, say where the fingers cross at 45-degree angles to the horizontal.

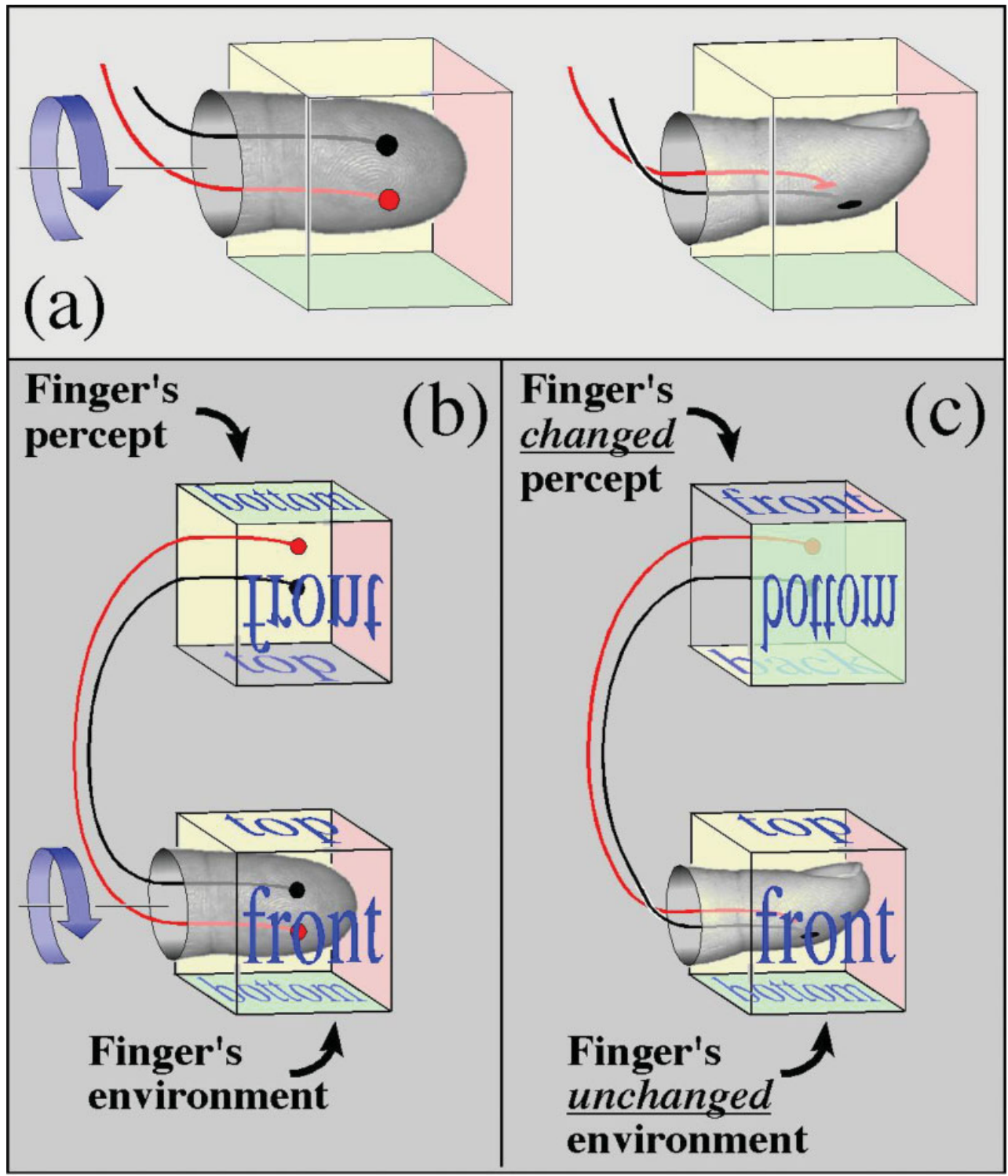

Fig. 3. Schematics of a finger, its environment, and its percept if connections are uncrossed. (**a**) Finger facing forward (left) and facing downward (right) in an unchanging environment. Sketched wiring and red/black spots are exactly those depicted in the simplified cartoon of Fig. 2b. Notice that when the finger rotates, it begins facing the transparent front of the environment and ends up facing the green bottom of the environment. (**b**) Detailed connections, environment, and percept of forward-facing arrangement. Notice that the percept of the lettering, "front," is flipped top-to-bottom, as necessitated by the reversal of the red and black spots in Fig. 2b. (**c**) Connections, environment, and percept of downward-facing arrangement. Notice that now the lettering, "bottom," is flipped upside-down, and "front" is now upright. Thus, as the finger is rotated, its percept changes even for a static environment. This would imply that interpreting Braille would be fundamentally different on the side versus the pad of a finger, and that somatotopic interpretation of the direction an insect walking on the skin would depend on its location and orientation. We propose that such a situation would be untenable for control or perception.

The clumsiness and loss of control flexibility is a mathematical consequence of the fact that the singularity shown in Fig. 2b is discrete: it occurs along lines separating upright from inverted regions of space. In the same way, the corners of the circumscribed square of Fig. 4 define discrete points, and a new, rotated square will define different points. It is not possible to develop a continuous control model from a discrete mapping such as this, and so flexibility is lost and:

1. each slightly different task imposes a new relearning process; this is a well studied issue, termed task specificity, in the neuroscience literature (Salter et al., 2004) and
2. to perform previously learned tasks requires some mechanism of looking up previously memorized tasks from a stored library, which is inevitably slower, clumsier, and more prone to error than executing a generalized and continuous algorithm.

Because these attributes are mathematically determined and follow only from geometrical constraints without reference to any particular physiology or organism, we conclude that any continuous 3D processing system that fails to cross connections and connects ipsilaterally through the reflection mechanism shown in Fig. 2b must suffer the specific control shortcomings described. This implies both that crossings play an important role in coordinated motion and that a lack of coordinated control must result if these crossings are not present.

To summarize, in example 1, the change in sign of the vertical coordinate in 3D mapping leads directly to a singularity that results in an inability to distinguish upright from inverted somatosensory images. This singularity can be removed by allowing connections to cross as in Fig. 1b: in this case, both the 3D "body" and the 3D "brain" become oriented identically. Similarly in example 2, a change in sign in the control problem of Fig. 4 generates singularities that interfere with motor function. This singularity can be removed by pointing both fingers in the same direction, which changes the minus sign directly (Bogaerts and Swinnen, 2001). Thus, the second example may be viewed as an analog (in physical space where we can see it) of the same structural concepts that are present in somatotopic network connectivity (which we cannot so easily see). In either case, we find that structural crossings have direct functional consequences.

## ALTERNATIVE CROSSING STRATEGIES

Because configurations of connections lacking crossings evidently present profound 3D sensory and motor control problems, we next turn to examining alternative configurations of crossings that, as we will show, have different quantitative and qualitative advantages over one another.

## Alternative 1: Rotation

A first alternative to the crossing scheme shown in Fig. 1b is illustrated in Fig. 5a, where the brain is depicted as being rotated about the $x$-axis (green arrows). We label top, bottom, and side surfaces of a body and a brain "cube" to clarify relative orientations. In this case, Fig. 5a (left) shows that the wiring of the connections A-A′ and B-B′ that appears in Fig. 1b is untangled as a result of turning the brain upside down. In this respect, the rotation is similar to the reflection that unravels crossings entirely in Fig. 2. On the other hand, although this crossing has been unraveled, a new set of crossings is introduced. This is shown in Fig. 5a (right), where we illustrate that the new points, C and D, which would not cross in the unrotated configuration (Fig. 1b), must now cross on their ways to their destinations, C′ and D′.

Rotation of the brain or body, therefore, trades one type of crossing for a second. All possible rotations will produce this result, because in the problem we have defined, connections between vertically displaced points cross, but those between horizontally displaced ones do not (cf. Fig. 1b). Any rotation that removes the crossings of vertically displaced points necessarily occurs about a horizontal axis, and such rotations produce new crossings of previously uncrossed connections [see (Fuller, 1978; Bauer et al., 1980) for mathematical explanations for this observation].

## Alternative 2: Decussation

A final approach to disentangling the crossed connections of Fig. 1b is to decussate connections across the midline as shown in Fig. 5b. In this state, the front face of the "brain" cube on the right would connect appropriately with the front face of the left "body" cube. Similar connections can be made between all the faces of the cubes. Here, the "brain" cube is separated into two lattices, one representing each side of the body. As shown in the figure, crossings appear in a lateral and decussated network, but they are between commissural tracts that traverse the midline, and not within tracts, or bundles, as before.

We have speculated in the Introduction that the decussated configuration presumably conveys some functional advantage over the alternatives shown in Figs. 1b or 5a. From our preceding analysis, it seems that crossings cannot be removed entirely without functional penalty, so we turn next to evaluate quantitatively how decussated crossings differ from other crossing geometries.

## ANALYSIS
### Numbers of Crossings and Somatotopy

Comparisons of the numbers of crossings associated with upright, rotated, and decussated configurations provide useful insights. In the upright configuration, connections between vertical columns of nodes naturally organize into bundles, as depicted in Fig. 6, inset (a). If there are $n_z$ vertical layers of nodes and $N = n_x n_y n_z$ nodes in all, then there must be $N/n_z$ bundles, each of which has $(n_z - 1) + (n_z - 2) + (n_z - 3) + \cdots + 1 = n_z(n_z - 1)/2$ crossings, so there are $N(n_z - 1)/2$ crossings in all. By contrast, as is shown in inset (b) to Fig. 6, in

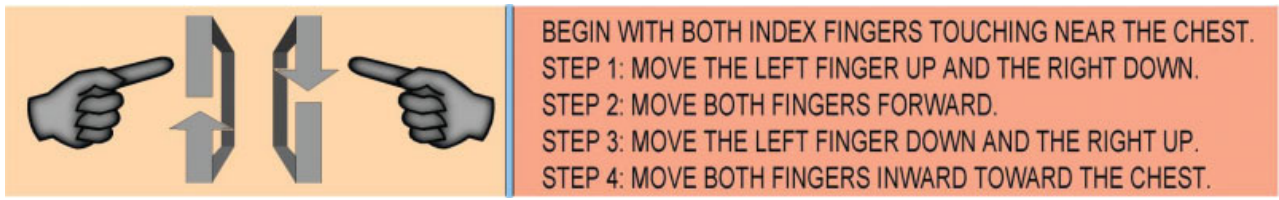

BEGIN WITH BOTH INDEX FINGERS TOUCHING NEAR THE CHEST.
STEP 1: MOVE THE LEFT FINGER UP AND THE RIGHT DOWN.
STEP 2: MOVE BOTH FINGERS FORWARD.
STEP 3: MOVE THE LEFT FINGER DOWN AND THE RIGHT UP.
STEP 4: MOVE BOTH FINGERS INWARD TOWARD THE CHEST.

Fig. 4.  Exercise illustrating the control difficulty when a normal and a reflected processing scheme coexist. This is an intrinsically 3D problem: the difficulty vanishes if both fingers are pointed forward.

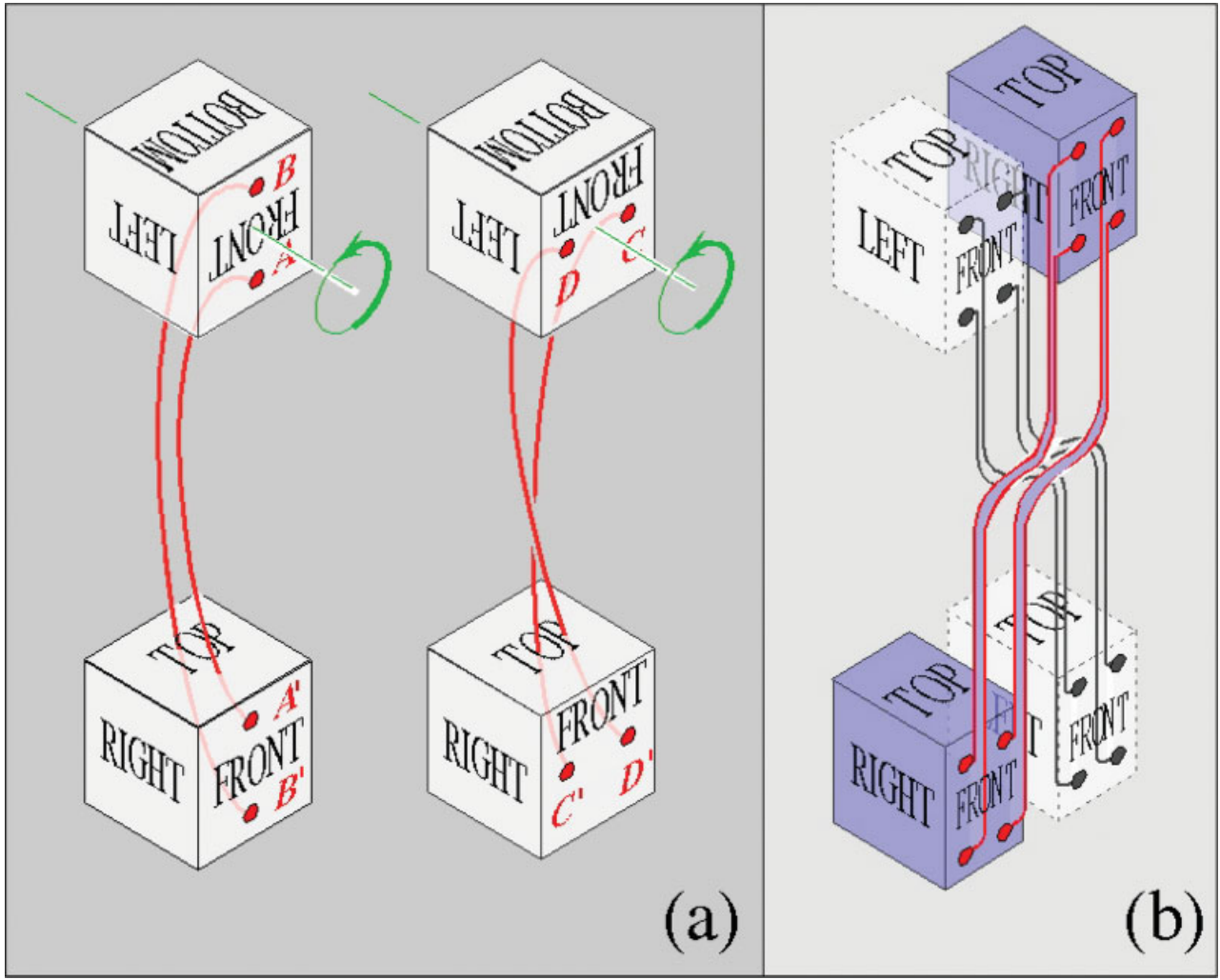

Fig. 5.  Cartoons of connections between 3D systems that are (**a**) rotated and (**b**) lateralized and decussated. (a, left) If the brain is rotated, for example, around the green axis shown, the original crossings indicated in Fig. 1 are removed, but (b) these are replaced with other crossings. (b) In a system divided laterally into halves, a decussated network can be established. In this case, crossings occur between tracts (shown in red and black), but there are no crossings within a tract as in other geometries.

the rotated configuration, nodes on the entire front surface connect via a single bundle that crosses at the midline, producing $n_z n_y(n_z n_y - 1)/2$ crossings. If there are $n_x$ such layers, then there are a total of $N(n_z n_y - 1)/2$ crossings in the rotated configuration.

This combinatorial arithmetic indicates that the **rotated configuration has more crossings than the** upright one. As each crossing must be navigated correctly during development to achieve an intended function, a configuration with many crossings is *prima facie* more prone to navigational, or "targeting," errors than **one with fewer crossings. From this standpoint, the** upright configuration is evidently superior to the rotated one. However, the rotated configuration has an advant-age over the upright configuration, because its connections are not broken up into separate bundles and so are intrinsically somatotopically ordered.

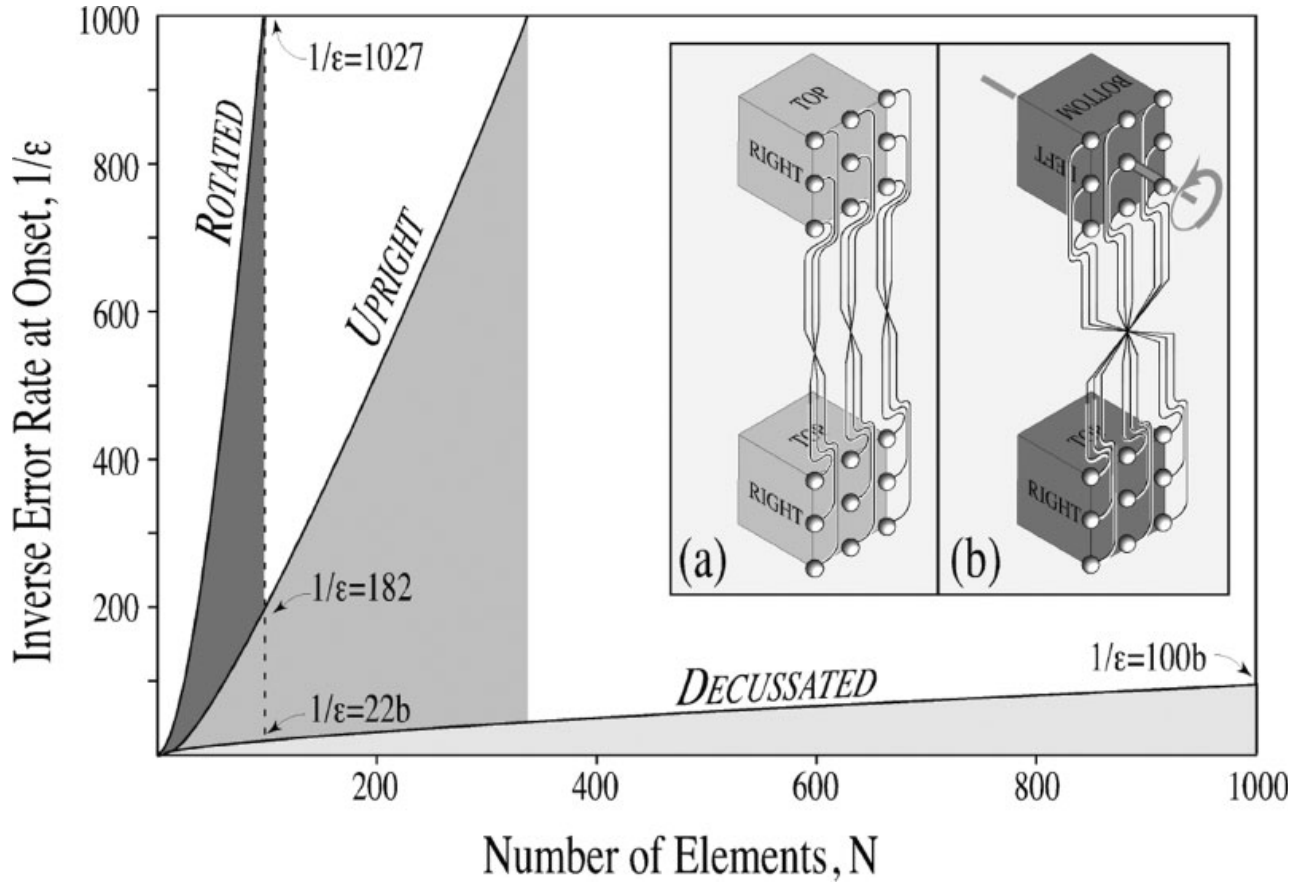

Fig. 6. Main plot: estimate of the relative robustness of three configurations to wiring errors as a function of the number of elements, N. As N grows, the error rate needed to prevent miswiring becomes dramatically lower for the decussated configuration than for either upright or rotated alternatives. Note that '*b*' is the number of bundles within a tract, as defined in the text. Insets: (**a**) in upright configuration, vertically separated nodes associate into tracts of crossed connections; (**b**) in rotated configuration, nodes cross one another within a single bundle (only the frontmost layer of lattice nodes are shown for simplicity).

In a well-ordered system, it is straightforward to compare the information required to establish somatotopic connections. If we calculate the number of times that somatotopic ordering is broken, we find that the rotated system has no changes in ordering from element-to-element but, in the upright configuration, the ordering of connections changes between each of the $N/n_z$ vertically oriented bundles. Thus, this system has $N/n_z - 1$ reorderings of connections, and correspondingly, at least $N/n_z - 1$ bits of information must be carried by any targeting or information processing procedure.

The third, or decussation, approach to disentangling the crossed connection essentially translates the lateral halves of brain or body into a decussated network as shown in Fig. 5b, by moving one-half of the brain (red connections) rightward as shown in Fig. 5b (or equivalently by inserting connections from the opposite side of brain nodes) and *vice versa* for the other half of the brain (black connections). This translation is remarkably effective in removing crossings. Notice that there are no crossings of connections at all within either colored tract in this configuration. As shown in the figure, crossings appear in a lateral and decussated network, but they are between commissural tracts that traverse the midline, and not within tracts, or bundles, as before.

In a decussated configuration, the number of neighboring connections that must negotiate crossings is at most the number of contacts between tracts (as connections within a tract remain adjacent to their neighbors). In a simple cubic brain or body, this number is $n_x^2$. This is an ideal bound, because decussation between lateral halves occurs via tracts of axons in real organisms. In such a case, the number of crossings is $bn_x^2$, where $b$ is the number of tracts present (Yoshikawa et al., 2003). In contrast with other ordering schemes, translations shown in Fig. 5b maintain exact somatotopic ordering except at the interface between left and right halves where the number of somatotopy violations is $n_x n_z$.

The information content required to accurately guide connections in a decussated configuration is $2b$, where $b$ is the number of bundles. Once a bundle is specified, a connection need only know **which side of the midline** to travel to and (provided it **engages in no additional deviations**) it will arrive correctly placed at its destina-

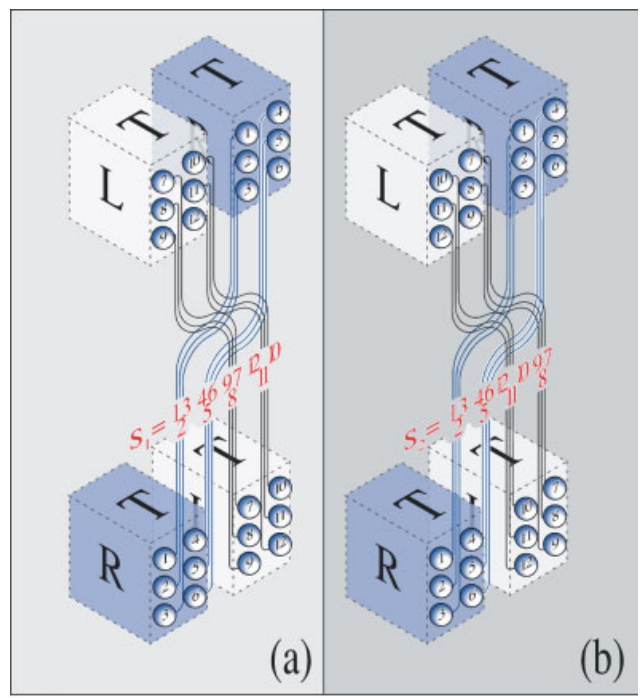

Fig. 7. Decussated wiring alternatives in lateralized system. (**a**) Laterally antisymmetric configuration with decussation across the midplane. (**b**) Alternate ordering of nodes: in configuration (a), numbering in the "body" is sequential, top-to-bottom and left-to-right, but this produces nonisotropic ordering of connections on one side. In (b) connections and nodes are all consistently and somatotopically ordered from lateral-to-medial.

### TABLE 1. Numbers of crossings of connections and violations of somatotopic mapping for the three configurations depicted in Fig. 1b ("upright"), Fig. 5a ("rotated"), and Fig. 5b ("decussated")

|  | Upright | Rotated | Decussated |
|---|---|---|---|
| Crossings | $\dfrac{N(n_z - 1)}{2}$ | $\dfrac{N(n_z n_y - 1)}{2}$ | $bn_x^2$ |
| Violations of somatotopy | $n_x(n_x - 1)$ | 0 | $n_x n_z$ |
| Required information content | $n_z$ | $n_y n_z$ | $2b$ |

tion parallels that of the somatotopic distribution of both motor and sensory functions. If lower numbers represent the more caudal parts of the body, and higher numbers represent the more rostral parts, the pattern of Fig 4b would suggest that the legs would be medially distributed, whereas the arms and head would be laterally distributed in the cortex. Both ordering schemes shown in Fig. 7 are mathematically permissible, and we speculate that the one that achieves better function in a given application (e.g., coarse vs. fine motor control) is selected for by evolutionary or developmental pressures.

## Effect of System Size

Thus far, we have explored three transformations that alter the character of crossings of connections in 3D. Although these examples are not comprehensive, they nevertheless present a clear means of quantitatively comparing categories of transformations. Table 1 summarizes the effects of rotation and decussation on numbers of crossings and violations of somatotopic ordering as a function of system size, as expressed as the number of nodes, $N = n_x n_y n_z$, in a cubic grid of dimensions $n_x$, $n_y$, and $n_z$ (we exclude the reflected case because of its functional irregularity).

Table 1 provides a concrete mechanism for evaluating relative fitnesses of the different configurations. Specifically, each crossing must be successfully navigated during development in order for axons to reach their correct destination (Changeux, 1985). Crossings, like any other mechanical operation, are inevitably associated with a finite error rate, particularly in physiological systems that are often prone to noise (Wiesenfeld and Moss., 1995; van Pelt et al., 1997). Therefore the number of times that a crossing will be erroneously navigated, and hence, that an axon will be diverted from its destined target, can be anticipated *a priori* to be simply proportional to the number of crossings.

In practice, systems that successfully operate in the presence of noise employ mechanisms for correcting errors. The CNS does this by producing many more neurons and connections than are necessary and then removing incorrectly connected fibers through axonal retraction or neuronal apoptosis (Gross, 2000). To explore the topological implications for error correction, in Table 1, we also itemize the number of times that adjacent connections deviate from regular somatotopic ordering, under the presumption that such

---

tion. Thus, a decussated network provides a marked reduction in the information required to establish somatotopy.

In systems that are lateralized into left and right decussating halves in this way, it is intriguing to note the following unexpected and subtle mathematical development. Consider a lattice of nodes as shown in Fig. 7, enumerated in the body top-to-bottom and then right-to-left. Any other systematic numbering scheme would work as well; this scheme is convenient for purposes of illustration. In this figure, we depict only the frontmost set of grid elements for simplicity. Enumerated in this way, connections in Fig. 7a become labeled: $S_1 = (1,2,3,4,5,6,9,8,7,12,11,10)$ as shown in red in the figure. That is, although the body is consistently ordered, one-half of the connections is somatotopic, but the other half is not. It is not clear whether there are physiological correlates to this mathematically allowable configuration; however, the existence of right- versus left-handedness suggests that motor function (and by inference, plausibly somatotopic organization as well) in opposing halves of an organism can differ significantly (Changeux, 1985).

An alternative ordering scheme that does not suffer from this somatotopic asymmetry is depicted in Fig. 7b: here brain and body are ordered from top-to-bottom and from lateral-to-medial. Now, the connections in either half remain somatotopically ordered, and only across the midline is there a discontinuity: the ordering here is $S_2 = (1,2,3,4,5,6,12,11,10,9,8,7)$. This organiza-ordering can only aid in directing neighboring connections to their correct targets.

From Table 1, it is evident that numbers of crossings grow faster than linearly with numbers of elements for upright and rotated configurations, but slower than linearly for the decussated case. This property produces profound differences in the numbers of elements at which crossing mistakes first appear. Figure 6 shows the effect of the number of elements, N, on inverse of the onset error rate, $1/\varepsilon$, at which miswirings can first be anticipated. **Miswiring can be expected below the error curve depicted in Fig. 6, but would be unlikely to occur above the error curve.**

Thus, in a system with N = 100 elements (broken vertical line in main plot), one can on an average tolerate an error rate of 1:22b in the decussated configuration before encountering miswirings in which a connection is diverted from its intended course by noise. As defined previously, $b$ is the number of bundles, and for Fig. 1, we assume all connections within a tract are in a single bundle ($b = 1$). In the upright and rotated configurations, by comparison, one needs to maintain an error rate of 1:182 and 1:1,028, respectively, to avoid miswirings. Naturally, from a design perspective, one would want to provide a safety factor, but these numbers give conservative estimates of the error rate at which miswirings are first to be expected.

From these estimates, for N = 100 the decussated network can produce an advantage of better than a factor of 8 in required error tolerance over the upright configuration, and nearly a factor of 50 over the rotated case. This advantage becomes more extreme as the number of elements (i.e., the system complexity) grows. Thus, in tiny networks of say 10 elements, the required error rates at which miswirings can first be expected are comparable for the three cases, whereas at N = 1,000 the contrast grows dramatically: miswirings will be found at $\varepsilon = 1:100b$, 1:4,500, and 1:49,500 in decussated, upright, and rotated configurations. Furthermore, from the final row in Table 1, it can be seen that for mammals with millions of axons within a single tract, the upright configuration requires hundreds of bits of information, the twisted configuration tens of thousands of bits, and the decussated network only twice as many bits as bundles crossing the midline. Thus, unless hundreds to many thousands of genes are dedicated to regulation of crossings during development, only the decussated arrangement can possibly be feasible.

This analysis implies that sufficiently large networks **are virtually required to decussate in order to reduce connection** errors if 3D somatotopy is needed. These results are generic: even if a system develops other strategies for reducing erroneous connections, for example, removal of erroneous connections or **decreased** numbers of bundles, for sufficiently large N, the decussated network will unavoidably **benefit from** fewer wiring errors.

## Evolutionary Simulation

Because miswiring of motor or sensory systems is likely to convey functional, and hence evolutionary, disadvantages, we speculate that there should be an evolutionary drive toward decussated networks in complex processing systems. Moreover, many proteins involved in developmental decussation, for example, netrins, slits, and UNCs, are highly evolutionarily conserved, suggesting that the biochemical pathways associated with these proteins have played a lasting and important role in evolutionary development.

On the other hand, as we have mentioned earlier, executing decussation successfully during development involves complicated and precise mechanisms, and so in very simple systems, for example, networks containing fewer than several hundred axons, error rates under 0.1% could be achieved in an ipsilateral, upright organization. Even the most primitive organisms such as *C. elegans* have several hundred neurons, and hence one would expect most nervous systems, particularly those that involve 3D spatial processing, to incorporate decussation as a mechanism for somatotopic error management.

Evidently, there is a dramatic quantitative advantage in robustness of complex decussated networks against occasional miswiring errors when compared with the alternative configurations considered here. Moreover, for the organization scheme shown in Fig. 5b, each lateral side of the network can be perfectly somatotopically organized, which we have suggested may also be of functional value. These mathematical results beg the question, however, of how the complex developmental mechanism of decussation, involving accurate pathfinding of potentially millions of axons through as many counter-directed trajectories, has evolved in so many different systems and organisms.

To investigate how this could occur through random mutations, we construct a direct network simulation that **randomly rewires itself and "evolves" so as to maximize its fitness** (Edelman, 1987; Koch and Laurent, 1999). Such a simulation is also of interest because it may provide insight into rewiring efficiency in regenerated systems. The simulation initially connects a 3D lattice of "brain" nodes point-to-point in the simplest way possible with an identical arrangement of "body" nodes in the same way as in the classic "Tea-trade" model (Willshaw and von der Malsburg, 1979; Dolman, 2002). The program then repeatedly reassigns connections by choosing two brain nodes at random and interchanging their connections. If the total number of crossings throughout the lattice network is reduced by the interchange, then it is kept, otherwise the original configuration is restored. A crossing is defined to occur if lines connecting the given nodes pass within a small distance, $\varepsilon$, of one another ($\varepsilon = 1/20$ of the internodal spacing in the simulations that we show). Notice that simulation selects for changes that reduce the number of crossings of axons anywhere in the network, but has no intrinsic preference for decussated or nondecussated configurations: whichever creates the smaller total number of crossings will be selected.

Figure 8 shows outcomes after 100,000 generations (i.e., random attempted interchanges), for two initial configurations of $10 \times 10 \times 10$ lattices of brain and body nodes. First, in Fig. 8a,b, we show **rewiring results** starting from a state in which the nodes were connected in the upright configuration of Fig. 1b. Second, in Fig. 8c,d, we show that the evolution starting from the decussated configuration of Fig. 5b. Thus, in the first, ipsilateral state, very few connections crossed a midplane to start with, whereas in the second, decussated configuration, almost all axons crossed the midplane in the Y-direction and almost none crossed the midplane in the X-direction. To simulate small wiring errors, each initial state is perturbed from this ideal state by switching the connections of 10 randomly chosen pairs (i.e. 2% of the 500 pair total).

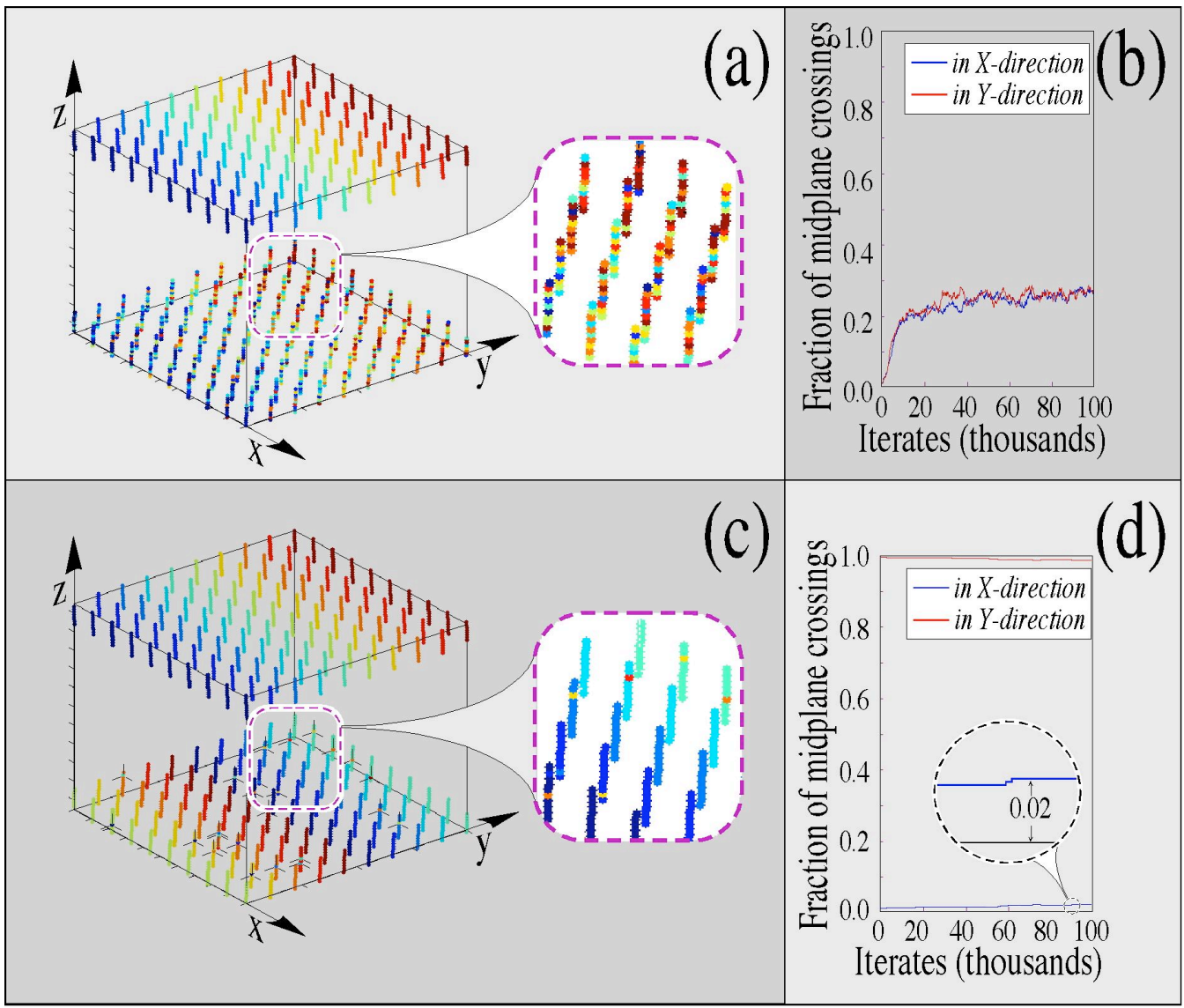

Fig. 8. Evolution of **(a)**, **(b)** nearly upright and **(c)**, **(d)** nearly decussated states described in text. Plots **(a)** and **(c)** show color coded correspondence between 1000 brain (above) and body (below) nodes after 100,000 generations of rewiring algorithm that reduces the number of crossings between neighboring connections, and **(b)** and **(d)** show associated fractions of midplane crossings in the two simulations. Note that in the decussated configuration, almost all connections cross the midplane (in the Y-direction), nevertheless the total number of crossings between individual connections is seldom improved by rearrangements.

To determine whether the prescribed random mutational rewiring results in more or less somatotopic ordering, in Fig's 6(a), we color code body nodes to agree with the color of the brain nodes to which they are connected. The brain node colors remain unchanged throughout the simulation – so initially the brain and body would look identical, but after 100,000 generations, the body becomes significantly disordered, as shown in the magnified view beside Fig. 8(a). To quantify the extent of ordering and decussation, in Fig. 8(b) we plot the fraction of connections that cross a midplane in either the X- or the Y-direction. For the upright state, midplane crossings grow until they saturate at about 1/4 crossings in either X or Y after about 10,000 generations. The 1/4 crossing state is a consequence of simple probabilistic considerations: once in this state, altering any connection has an equal probability of improving or worsening the number of crossings. In short, the simulation shows that mutational rewiring of an ipsilateral network increases the number of intra-tract crossings, saturating with about a quarter of the axons crossing the midplane. The rewiring also results in a degradation of somatotopic mapping.

This situation is quite different if the system is initially decussated. In Fig. 8(c)-(d) we show the corresponding evolution starting from a state that is initially decussated in the Y-direction and ordered in the X-direction, but with 10 pairs of randomly chosen connections interchanged as before. We identify the disordered connections after 100,000 generations in Fig. 8(c) with black tripods. The magnified inset illustrates that the initial ordering is largely retained: anomalously colored dots indicate the few nonsomatotopic connections. In the final state, 2.5% of connections are disordered when compared with 2% initially. If we quantify the midplane crossings as before, we find this tendency to remain ordered is confirmed, as shown in Fig. 8d.

These simulations indicate that random mutations have limited ability to reduce the numbers of crossings. Specifically, the simulations demonstrate that there is a dramatic quantitative advantage in robustness of decussated networks against miswiring events when compared with the alternative ipsilateral configuration. If an ipsilateral 3D Cartesian network of 1,000 nodes randoml rewires to reduce crossings of connections, then an initial small fraction of wiring errors will diverge to approach a stochastically dominated state within 10,000 generations. In contrast, a decussated configuration with the same initial wiring error will remain stable over 100,000 generations. Associating connections into tracts or bundles (not done in this simulation) would only accentuate the advantages of the decussated configuration, as indicated in our earlier analysis of the effects of system size.

## DISCUSSION
### Mathematical Findings

This study gives rise to several unambiguous mathematical findings. Some of these findings are counterintuitive; nevertheless they are matters of mathematical certainty independent of physiological considerations. In particular the following findings apply to all functional lateralized, point-to-point, somatotopic, 3D networks:

1. Crossings must occur as a matter of mathematical necessity, regardless of network complexity. In simpler or more primitive systems, these crossings may occur within a tract (as in Fig. 1b). In more complex systems, however, for reasons of reliability, crossings are strongly favored outside of tracts through the formation of decussated pathways (as in Fig. 5b).
2. There is a maximum number of neurons (between 100 and 500 depending on assumed targeting error rates) above which only a decussated configuration can be reliably wired. 3D systems larger than this maximum are unstable to rewiring events if wired ipsilaterally, but are significantly more stable if decussated across the midline to the contralateral side.
3. If 3D motor or sensory systems were nevertheless wired without crossings (as in Fig. 2a), function will be compromised due to the need to process coexisting but contradictory spatial information that changes with orientation of the body.
4. Two identifiable wiring states appear in decussated arrangements. In one state, half of the body is somatotopically wired and the other half is not, implying better coordination on one side of the body than on the other. In a second state, wiring is symmetric left-to-right, but is asymmetric lateral-to-medial. These two states can coexist in different systems, for example, the gross motor system can be of one type and the fine motor control can be of the other.
5. In systems with millions of connections, either the decussated mechanism of crossing must be adopted,

or hundreds to thousands of genes must be dedicated to the regulation of crossings during development.
6. Finally, it is notable that only one of brain or body need fold for the considerations that we address to come into play. Thus, in Fig. 1d, we show that folding only the body in the manner done in Fig. 1b produces no crossings, whereas in Fig. 2c, we show that folding only the body in the manner of Fig. 1b does produce a crossing.

### Physiological Interpretation

These findings have a number of physiological implications. First, the findings suggest that decussated pathways are prevalent in vertebrate nervous systems because decussation minimizes pathfinding errors and genetic information content required, and thereby provides significant evolutionary advantages. Surprisingly, even in small networks of only 100–1,000 neurons, on the order of the number of neural connections in common insect tracts, decussation can reduce the rate of erroneous connections by over an order of magnitude. Hence, we predict that 3D networks larger than this, and so spatially dependent tracts of animals only as complex as insects, must decussate in order to function reliably. It is interesting to note in this context that many proteins involved in developmental decussation, for example, netrins, slits, and UNCs, are highly evolutionarily conserved and are seen in the most primitive of insects, worms, and other invertebrates as well as in vertebrates. The presence of an elementary decussated tract (the ventral tract) even in *C. Elegans* suggests, indeed, that the evolutionary advantage of decussation may have predated the creation of complex visual organs or limbs, contrary to the dominant paradigm described in (Vulliemoz et al., 2005).

Second, the findings may have implications for the regeneration of somatotopic connections following injury. Major decussated systems (e.g., the corticospinal and spinothalamic tracts) are often disrupted by spinal cord injury (SCI). Current neuronal regeneration strategies are based on the hope and expectation that if spinal axons can be regrown, motor and sensory function will be restored. Our results show that functional recovery will depend on specific and appropriate crossing pathways being maintained. Thus, for example, if regeneration occurs toward an upright (Fig. 1b) or an inverted (Fig. 2) state with equal likelihood (and we know no *a priori* reason to assume otherwise), then fully half of patients can be expected to suffer functional impairment associated with processing coexisting but contradictory inputs.

The indication that a lack of crossings may cause somatotopic motor and sensory dysfunctions carries particular urgency in regeneration research, where it has been demonstrated that CNS axons can be provoked to sprout across the midline (Schnell et al., 1994) following SCI, thus undoing the crossings accomplished by decussating tracts. Thus our findings imply that is imperative that contralateral, decussated, connections be maintained; otherwise, somatotopic sensation and coordinated motion will be defeated.

Third, the findings potentially give insight into developmental disorders. Several neurological dysfunctions are associated with the failure of spinal tracts to decussate during development (Cremer et al., 1994; Jouet et al., 1994; Vits et al., 1994; Wong et al., 1995; Cohen et al., 1998). It is known that decussation is mediated during normal development by cellular adhesion molecules (CAMs) including the L1 subfamily (Grumet et al., 1991; Cohen et al., 1998; Fitzli et al., 2000; Lustig et al., 2001; Backer et al., 2002), and the failure to properly express these molecules has been linked to developmental disabilities including deficits in spatial learning (Jouet et al., 1994; Vits et al., 1994; Wong et al., 1995). Additionally, it has been demonstrated that disrupting the expression of these CAMs during development interferes with pyramidal decussation in a mouse model, producing living animals with motor and spatial learning deficits (Cremer et al., 1994; Cohen et al., 1998). Similar results are known in humans as well, where mutations of L1 produce developmental disabilities (SPG1 and MASA syndrome) that involve deficits in spatial learning (Cremer et al., 1994; Jouet et al., 1994; Vits et al., 1994; Ferland et al., 2004). Thus, one of the chief regulators of decussation is also implicated in motor and spatial learning functions, seemingly in the way defined by our geometric analysis. Likewise, in Joubert syndrome, a congenital failure to decussate is believed to be a significant contributor to severe clumsiness in afflicted patients (Ferland et al., 2004).

These data are not definitive, but are consistent with **our mathematical predictions that networks that have no intratract crossings and in which decussation has been defeated (e.g., Fig. 2a) must give rise to spatial dysfunctions.** We look to future investigations to identify which of these potential links between structure and function **are real and which may be merely coincidental.**

## Simplicity Meets Reality: Comparisons

Obviously the mammalian CNS is nowhere as simple as the caricatures sketched in our figures. The CNS contains nondecussated, partially decussated, decussated, and even doubly decussated (Pettigrew, 1991) pathways. Moreover, several motor and sensory tracts have multiple stages, only one of which decussates. And finally, even some ancient tracts partially decussate. Therefore, appropriate care needs to be taken in comparing specific neuronal structures with the mathematical findings and physiological interpretations that we have outlined. A crude inventory can be made by comparing the human gross and fine motor tracts, because gross control involves fewer connections than fine control. Here, as predicted from our model, almost all gross motor tracts (which require comparatively few connections) are ipsilaterally wired. This includes the lateral vestibulospinal tract, the medial reticulospinal tract, part of the medial vestibulospinal tract, and most of the medial corticospinal tract. The only exception is the tectospinal tract, and this notably coordinates head movement in response to 3D stimulae. By contrast, almost all fine motor tracts (requiring comparatively more connections) are decussated. These include the lateral corticospinal tract and the rubrospinal tract. An exception is the lateral reticulospinal tract, only part of which decussates.

Exceptions such as this reinforce the importance of due care in interpreting our results, and lead us to examine three illustrative examples in detail. In all cases that we are aware of, even when a particular subsystem seems to operate without decussating, detailed examination confirms that complex 3D somatotopic function is invariably governed by decussated pathways, whereas simpler or 2D systems typically do not decussate.

A first example that illustrates this point is the rubrospinal tract: this is a nondecussating motor pathway that projects ipsilaterally from the red nucleus to specific spinal motoneuron pools (Kuchler et al., 2002). The corticorubral projection does decussate, however, and is somatotopic in the macaque (Burman et al., 2000). Thus, although decussation does not appear in the form seen in the classic pyramidal level of the medulla, nevertheless decussation has evolved, in a different way, in the rubrospinal system. By comparison, the dorsal column pathway ascends ipsilaterally to terminate in the gracilis and cuneatus nuclei in the brainstem, and the axons of these nuclei then project somatotopically to the contralateral thalamus [e.g. in the rat (Levinsson et al., 2002; Shaw and Mitrofanis, 2002), the mole (Catania and Kaas, 1997), and the raccoon (Rasmusson and Northgrave, 1997)]. The presence of these many morphologically and phylogenetically different decussation architectures in different 3D somatotopic functional systems reinforces our assertion that decussation is invariably selected for because it plays an essential role in these systems.

A second example is important to recovery of locomotor function following SCI. Adult rats with hemisected spinal cords have been found to recover bilateral locomotion within 2 weeks after lesioning (Schnell et al., 1994; Tatagiba et al., 1997). This recovery has been shown to be associated with sprouting of axons across the midline from the intact side (Saruhashi et al., 1996; Thallmair et al., 1998). Because the new connections cross the midline, this example could appear to call the significance of decussation into question. However, 3D somatotopic connection through the injury site is not required for locomotor function because this function is coordinated by a central pattern generator (CPG) present in the L2 spinal cord in rodents (Nishimaru and Kudo, 2000; Butt et al., 2002) and humans (Dimitrijevic et al., 1998; Herman et al., 2002). As 3D somatotopy is not required to activate the CPG, either a decussated or an ipsilateral connection would work equally well from our analysis.

A third example concerns the only direct motor projection from the cortex to the spinal cord: the corticospinal tract. Here, on the one hand, cortical axons that project somatotopically to the spinal cord decussate in primates (Wu et al., 2000) and in the rat (Rivero-Melian, 1996). On the other hand, the rat does have a small ventral nondecussating pathway that can contribute to locomotor recovery (Loy et al., 2002), and so the role of decussation in locomotion itself may be questioned. Data supporting this conclusion include recent results reporting that rats are able to walk with 5% of ventral spinal white matter (You et al., 2003). Importantly, however, the ventral pathway does not contribute to skilled forepaw movements: in fact, these 3D somatotopic functions are lost when the decussated tract is cut (Whishaw and Metz, 2002). These results support our analyses indicating that the pathways that decussate are precisely those that rely on effective function of spatially dependent and phylogenetically advanced systems.

To summarize, available anatomical and functional evidence is consistent with our mathematical findings that decussation is essential to complex 3D somatotopic function. In our first example, we illustrated the fact that decussation has been selected for in different ways for different 3D spatially dependent systems, but it is invariably selected for nonetheless. In our second example, we emphasized that nonspatially dependent functions can be activated with either a decussated or a nondecussated pathway. In our third example, we summarized data indicating that it is the spatially dependent and phylogenetically advanced functions that rely on decussated pathways. Thus, although there may well be more than one influence leading to the evolution of decussated pathways, it appears from our analysis that the need to preserve 3D somotopy may play an important role. In the end, when describing the nervous system the best that can be done is perhaps to quote Oscar Wilde (Wilde, 2003), "The truth is rarely pure and never simple."

## CONCLUSIONS

The relation between decussation and spatial motor and sensory function is especially important in SCI. SCI causes massive reorganization of neuronal projection in rats (Jain et al., 1998), monkeys (Florence and Kaas, 1995; Jain et al., 2000), and humans (Moore et al., 2000), and several well-known groups are investigating strategies to promote the regeneration of spinal axons across damage sites. Yet it is not at all clear whether once successful strategies are developed, regenerated spinal axons will engage in specific reconnection to appropriate targets. Specific reconnection does occur in the peripheral nervous system and in the neonatal CNS. For example, crushed peripheral nerves maintain guidance cues for axonal regrowth and retrace their original path including branch points (Nguyen et al., 2002), and neonatal rats that regenerate across a thoracic cord transection site can recover locomotion with forelimb-hindlimb coordination (Hase et al., 2002). In adults, however, the case is considerably less certain: IN-1 treatment regenerates the corticospinal tract and restores locomotor function but as we have discussed, the mechanism appears to be due to sprouting of uncut corticospinal (Raineteau et al., 1999) and rubrospinal tract axons (Raineteau et al., 2002). Because this sprouting crosses the midline (Thallmair et al., 1998), it seems arguable that even the most elementary guidance mechanisms (which would present such crossings) may not be active in the adult CNS. In light of our findings, it seems imperative to establish (a) the extent to which SCI reconnection after injury is specific and (b) the capacity of the CNS to adapt to incorrectly rewired axons. Until these questions are definitively addressed, it will not be possible to predict the outcome of regenerative therapies for SCI patients, even once regrowth of CNS axons is effectively established.

## ACKNOWLEDGMENTS

The authors thank Bernard Coleman, Richard Nowakowski, and Kathryn Scarbrough for valuable insights and informative discussions.